\documentclass[aps,prb,twocolumn,groupedaddress,showpacs]{revtex4}
\usepackage{epsfig}
\usepackage[dvipsnames,usenames]{color}
\usepackage{color}
\usepackage{amsmath}

\begin{document}

\title{Density of States and Magnetic Correlations at a
Metal-Mott Insulator Interface}

\author{M.~Jiang$^{1,2}$, G.G.~Batrouni,$^{3,4}$ and R.T.~Scalettar$^1$}

\affiliation{$^1$Physics Department, University of California, Davis,
  California 95616, USA}
\affiliation{$^2$Department of Mathematics, University of California, Davis,
  California 95616, USA}
\affiliation{$^3$INLN, Universit\'e de Nice-Sophia Antipolis, CNRS;
  1361 route des Lucioles, 06560 Valbonne, France}
\affiliation{$^4$Institut Universitaire de France}

\begin{abstract}
  The possibility of novel behavior at interfaces between strongly and
weakly correlated materials has come under increased study recently.  In
this paper, we use determinant Quantum Monte Carlo to determine the
inter-penetration of metallic and Mott insulator physics across an
interface in the two dimensional Hubbard Hamiltonian.  We quantify the
behavior of the density of states at the Fermi level and the short and
long range antiferromagnetism as functions of the distance from the
interface and with different interaction strength, temperature and hopping
across the interface.  Induced metallic behavior into the insulator is
evident over several lattice spacings, whereas antiferromagnetic
correlations remain small on the metallic side.  At large interface
hopping, singlets form between the two boundary layers, shielding the
two systems from each other.
\end{abstract}

\maketitle

\section{Introduction}

Over the last decade, advances in synthesis techniques have made
possible the construction of well-defined interfaces involving
strongly correlated materials, notably transition metal
oxides~\cite{Zubko11,Mannhart10,Munakata11,Freeland10}.  Indeed, the
early discovery that the interface between two insulating oxides,
LaAlO$_{3}$ and SrTiO$_{3}$ is a high-mobility two dimensional
conductor~\cite{Ohtomo02} or even a superconductor~\cite{Thiel06} has
emphasized that novel physics can arise at such boundaries, beyond a
simple interpolation between materials properties on either side.
Subsequently, numerous heterostructure interfaces have been shown to
exhibit unique phenomena that are not present in their bulk
constituents.  In particular, Munakata {\it et al.}~presented a
subtle proximity effect that arises between a normal metal and an
antiferromagnetic Mott insulator, which can be understood in the
framework of Ruderman-Kittel-Kasuya-Yoshida (RKKY)
interactions~\cite{Munakata11}.

Advances in characterization techniques are also expanding the horizon
of the field.  It was shown recently that, by setting up a standing
wave of the incident X-rays and adjusting the position where the
intensity is peaked inside the sample, one can measure electronic
excitations as a function of location in the sample.  This standing wave
angle resolved photoemission spectroscopy (SW-ARPES) \cite{gray10}
method allows a depth selective probe of buried layers and interfaces,
and hence the construction of chemical and electronic structure
profiles within the sample.

Various numerical methods have been employed to model qualitative features of these experiments.  
Early attempts include the study of LaAlO$_{3}$ and
SrTiO$_{3}$ systems in which the electric fields arising from
the La$^{3+}$/Sr$^{2+}$ charge difference and the on-site interactions
are treated within the Hartree-Fock approximation~\cite{Okamoto}. 
A metallic interface between band and Mott insulators in
a quasi-one-dimensional lattice was explored with the Lanczos
method~\cite{kancharla06}.  Important insights into electronic properties at
interfaces have also been gained by dynamical mean field theory
(DMFT)~\cite{OkamotoDMFT}.  For example, it has been suggested that the Kondo effect
governing the interface between metal and Mott insulator is
inefficient so that Mott insulators are
impenetrable to the metal~\cite{helmes08}.  More specifically, the
quasiparticle weight decays as $1/x^2$ with distance $x$ from the
metal, but the prefactor of this decay was found to be very small.
Zenia {\it et al.}, however, emphasized that even a small proximity
effect can induce density of states to open up a metallic channel
inside an insulator sandwiched between to metals at sufficiently low
temperatures~\cite{zenia09}, leading to perfect conductance.  This
Fermi liquid is ``fragile":  finite temperature, disorder, or
frequency rapidly return the behavior to that of a conventional N-I-N
junction.  Possible device applications were suggested as a
consequence of this
sensitivity.~\cite{Zubko11,Mannhart10,Munakata11,Freeland10}
In a related work~\cite{Gutzwiller10},
a Gutzwiller approximation approach
was extended to inhomogeneous systems.  The decay length of the
exponential fall-off of the penetration of metallic character into the
insulator region was shown to diverge as the metal-insulator
transition is approached.

A further interesting set of studies involves the dynamical response
of strongly correlated systems with an interface, for example the
possibility of the suppression of charge transport, current
rectification, and the behavior of holon-doublon pairs, which are
important for devices \cite{silva10}.  So far, such dynamic phenomena
have been explored primarily for one-dimensional systems, where
time-dependent Hartree Fock~\cite{yonemitsu07} and density-matrix
renormalization group (DMRG) methods are especially
effective~\cite{white04}.

In this paper, we address the metal-insulator interface problem by
using determinant Quantum Monte Carlo
(DQMC)\cite{blankenbecler81,white89}, a numerically exact
approach, to treat correlated electron models.  Here we focus on an
inhomogeneous Hubbard model on a single two dimensional lattice
divided by a linear interface in two regions, one weakly correlated
(with $U=0$), and one at intermediate coupling (with $U$ non-zero).
The chemical potential is chosen to maintain particle-hole symmetry,
so that all sites of the lattice are on average half-filled.  This
choice avoids the fermion sign problem~\cite{loh90} and enables the
evaluation of magnetic correlations and the density of states at low
temperature.

Such DQMC simulations are at present limited to lattices of $400-1000$
sites (depending on the interaction strength and temperature).  By
focussing on a linear interface in a 2D lattice, we are able to
explore systems with a fairly large linear extent.  Most of our
results will be for $20\times20$ lattices.  This enables us, for example, to
evaluate the penetration depth across the boundary, since the
interface effects have sufficient room to heal before the lattice
edge.  The behavior of the Hubbard model across a planar interface in
a 3D material on systems of smaller linear extent has been explored by
Euverte {\it et al}~\cite{euverte12}.  The key finding is that up to
an interface hopping $V$ which is on the order of the bulk
hybridization $t$, the effects of the interface can extend well past
the two layers immediately at the interface.  That is, the interface
affects properties $3-4$ layers deep on both the strongly and weakly
correlated sides of the boundary.  When $V \geq 2t$, however, there is
a return to the values characteristic of decoupled materials with
$V=0$.  This ``revival" of magnetic order on the insulating side is
driven by the formation of singlets between fermions on the two layers
immediately adjacent to the interface which acts
to decouple the two materials.

This paper is organized as follows: In section II we explicitly write
down the Hamiltonian and provide a very brief overview of DQMC.  Section
III focuses on the magnetic properties of the 2D lattice with a
linear interface, both the near-neighbor spin correlation and the
structure factor.  Here the interaction strength $U$ and temperature
$T$ are varied at fixed interface hopping $V$.  The key result is a
determination of the penetration depth as a function of $T$ and $U$.
Section IV extends these results to the density of states.  Section V
examines the variation with $V$, and Section VI summarizes our
results.

\section{Model and methodology}

We consider the two-dimensional Hubbard model
\begin{equation}
\begin{split}
    \hat{H} = &-t \sum\limits_{\langle ij \rangle,L,\sigma} (c^{\dagger}_{iL\sigma}c_{jL\sigma}^{\vphantom{dagger}}+h.c.) - \mu \sum\limits_{i,L,\sigma}  n_{iL\sigma}\\
        &- \sum\limits_{i,\langle LL' \rangle,\sigma} t_{LL'} (c^{\dagger}_{iL\sigma}c_{iL'\sigma}^{\vphantom{dagger}}+h.c.)\\
        &+ \sum\limits_{i,L} U_{L} (n_{iL\uparrow}-\frac{1}{2}) (n_{iL\downarrow}-\frac{1}{2})
\end{split}
\end{equation}\label{Hubbard}

\begin{figure}
\psfig{figure=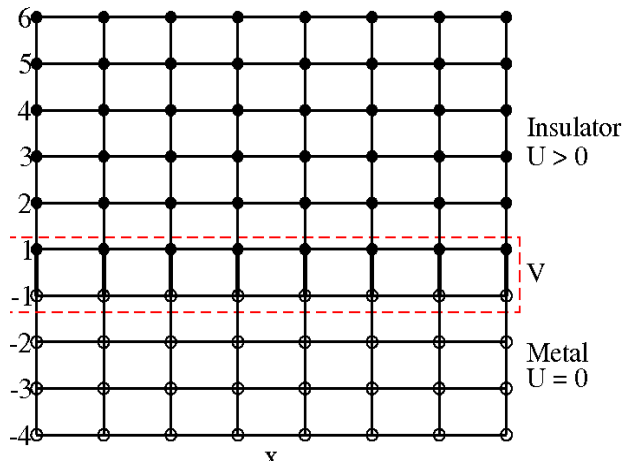,height=7.0cm,width=9.0cm,angle=0,clip}\\
\vskip -0.0cm
\caption{(color online) Geometry of the two-dimensional
    lattice with a one-dimensional interface (dashed red box).  The metallic
    lines (negative L) have $U=0$ while the insulating lines (positive
    L) have $U\neq 0$.  The lattice has periodic boundary conditions in
    the $x$ direction and open boundary conditions in the $L$
    direction. $V$ denotes the hopping/hybridization (heavy lines)
    across the interface. We adopt the lattice size $N=20\times 20$
    throughout the paper, unless otherwise indicated.}
\label{geom}
\end{figure}

Here $c^{\dagger}_{iL\sigma}(c^{\phantom{\dagger}}_{iL\sigma})$ are
fermionic creation (destruction) operators of spin $\sigma$ at site
$i$ in line $L$.  While $i,L$ can also be regarded as the $x$ and $y$
site labels $i_x,i_y$, the current notation emphasizes the broken
rotational symmetry (the interface is parallel to the $x$ axis).  Our
labeling convention is $L=\cdots -3, -2, -1$ for the $U=0$ lines and
$L=1, 2, 3, \cdots$ for the $U\neq 0$ lines, so that the interface
connects $L=\pm 1$.  $t$ and $t_{LL'}$ are the intra and inter-line
nearest-neighbor hoppings.  $t_{LL'} = t$ except at the interface
where $t_{-1,1} = V$. The geometry is shown in Fig. \ref{geom}.

The DQMC approach provides an exact solution to the finite temperature
properties of $H$ by decoupling the interaction term $U$ through the
introduction of an auxiliary Hubbard-Stratonovich (HS) field.  The
resulting action is quadratic in the fermion operators, so that the
trace over those coordinates can be performed, leaving determinants
(one for spin up and one for spin down) of matrices whose dimensions are
the spatial lattice size, and whose entries depend on the HS field.
The auxiliary field is sampled stochastically, and the values of the
up and down spin fermion Green's function for the configurations
generated are used to construct the various observables.  The CPU time
for the algorithm scales as the cube of the lattice size and
linearly \cite{footnote3}
with inverse temperature $\beta$.

At low temperatures, the fermion determinants can become negative,
precluding their use as a probability for sampling the HS field.  To
avoid this ``sign problem", we consider the half-filled case, $\mu=0$.
Here particle-hole symmetry (PHS) implies that $\langle
n_{iL\sigma}\rangle=0.5$ for all temperatures and interaction
strengths, even if $U_L$ vary spatially with $L$ and the hybridizations
are not all equal.  PHS can also be
used to demonstrate that the up and down spin determinants, although
they can individually become negative, always have the same sign.  As
a consequence, their product is always positive, allowing the study of
the low temperature behavior at the metal-insulator interface.

\begin{figure}
\vskip 0.3cm
\psfig{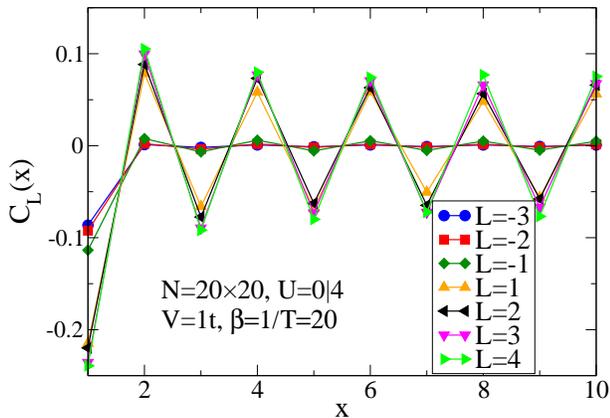}\\
\caption{(color online) Spin correlation $C_{L}(x)=\langle S_{i,L}
  S_{i+x,L} \rangle$ as a function of separation $x$ for different
  lines $L$.  Diminishing antiferromagnetism of the insulator adjacent
  to the interface is shown. The induced antiferromagnetic long-range
  orders in the metallic region are evident in spite of their
  weaknesses. $U=0|4$ denotes the interface setup.}
\label{clU4B20}
\end{figure}

\section{Magnetic properties}

We gain quantitative information concerning the spin correlations in
different lines by measuring the spin correlation $C_{L}(x)=\langle
S^-_{i,L} S^+_{i+x,L} \rangle$, with $S^-_{i,L} =
c^{\phantom{\dagger}}_{i \uparrow} c^{\dagger}_{i \downarrow}$,
parallel to the interface.  Figure \ref{clU4B20} shows that
antiferromagnetism in the insulating layer $L=1$ immediately adjacent
to the metal is diminished by 20-30\% relative to larger positive $L$
which are deeper in the Mott insulator.  Data are shown for boundary
hybridization $V=t$ and interaction $U=4t$ in the insulator.  Despite
a somewhat smaller amplitude, the correlations in $L=1$ appear to
remain long ranged.  Induced antiferromagnetic long-range
order\cite{footnote1} in the metallic region is also evident, although
its amplitude is an order of magnitude smaller than that on the
insulating side.

\begin{figure}
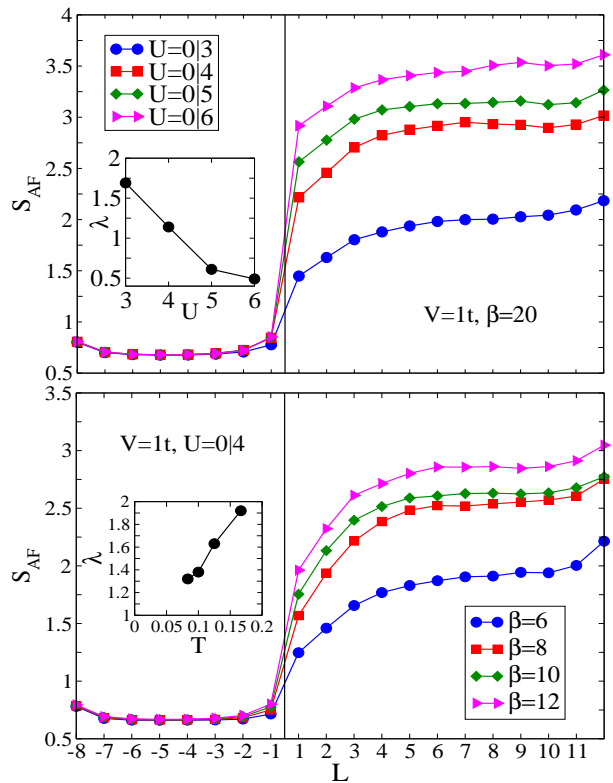

\psfig{figure=SafB20.eps,height=5.0cm,width=8.0cm,angle=0,clip}\\
\psfig{figure=SafU4vsT.eps,height=5.3cm,width=8.0cm,angle=0,clip}\\
\caption{(color online) AF structure factor $S_{AF}(L)$ in
each line $L$ for different correlation strengths $U$ and temperatures
$T$. The smooth penetration of
antiferromagnetic order into the metal and the diminishing AF order in
the insulator are clearly seen.
Fitting $S_{AF}(L)$ to a
hyperbolic function~\cite{openboundary},
$a\tanh L/\lambda+b$ allows the extraction of a penetration length
$\lambda$, shown in the insets.  $\lambda$ decreases with larger
correlation $U$ and lower temperature $T$. Further information}
\label{SafU4vsT}
\end{figure}

Compact information on the antiferromagnetic long-range order for each
line $L$ can be obtained by measuring the antiferromagnetic (AF)
structure factor,
\begin{equation}
   S_{AF}(L)=\frac{1}{N} \sum\limits_{x} (-1)^{x}\cdot C_{L}(x)
\label{saf}
\end{equation}
where $N$ is the linear lattice size.  Figure~\ref{SafU4vsT}
illustrates the $L$-dependent AF structure factor for different
correlation strengths $U$ and temperatures $T$. The smooth penetration
of antiferromagnetic order into the metal and the diminishing AF order
in the insulator are clearly seen. As might be expected, the induced
long range AF order in the metal adjacent to the interface is stronger
for correlated insulator with larger\cite{footnote4} on-site repulsion $U$
(Fig.~\ref{SafU4vsT}a) and/or lower temperature
(Fig.~\ref{SafU4vsT}b).  The overall magnitude of $S_{AF}$ in the
metal is rather small, consistent with the small real-space
correlations in Fig.~\ref{clU4B20}.  The results of
Figs.~\ref{clU4B20},\ref{SafU4vsT} are rather different from
previous work of Sherman {\it et. al.}
\cite{Sherman} which found that the antiferromagnetic
order can penetrate into the metal to a depth of ten lattice spacings,
but are consistent with the shorter range effects described in
\cite{helmes08,zenia09}.  Figure~\ref{SafU4vsT} indicates that by
layer $L=-3$ the influence of the contact with the insulator is
minimal.

On the other hand, we find that contact with the metal has a
substantially greater effect on the insulator.  Here the ``Kondo
proximity effect" diminishes the long range AF order of the insulating
lines adjacent to the interface.  This tendency to paramagnetism
competes with the long range order induced by the exchange energy $J
\sim t^2/U$ deep in the insulator (Fig.~\ref{SafU4vsT}a).  To quantify
the proximity effects, we fit the curves of $S_{AF}(L,U,T)$ with the
hyperbolic form $S_{AF}(L,U,T)= a\tanh L/\lambda+b$. The penetration
depth $\lambda$ decreases with larger correlation $U$.  Interestingly,
$\lambda$ increases at higher temperature.  It is possible
that raising $T$ weakens the magnetic order in the insulator, thereby
enhancing the effects of the contact with the metal.

\begin{figure}
\vskip 0.3cm
\psfig{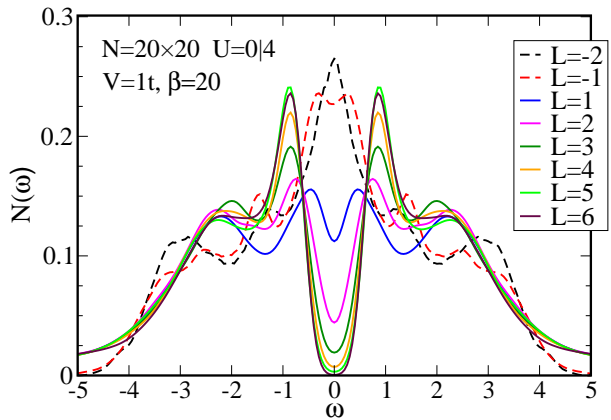}\\
\caption{(color online) The $L$ dependence of the density of states
  $N(\omega)$ shows the evolution of the spectral properties across
  the metal-insulator interface. Lines of sites within the Mott
  insulator ($L>0$ have nonzero $U$) are characterized by the presence
  of Kondo resonance peaks split by antiferromagnetic order at
  half-filling.  However this AF gap between the two peaks is
  dramatically weakened as $L \rightarrow 1$.  On the other hand, the
  densities of states associated with lines of sites on the metallic
  side of the interface show little effect of contact with the
  insulator.  Even for the line $L=-1$ most immediately adjacent to
  the boundary there is only a small dimple in $N(\omega)$ at
  $\omega=0$.}
\label{Axel}
\end{figure}

\begin{figure}
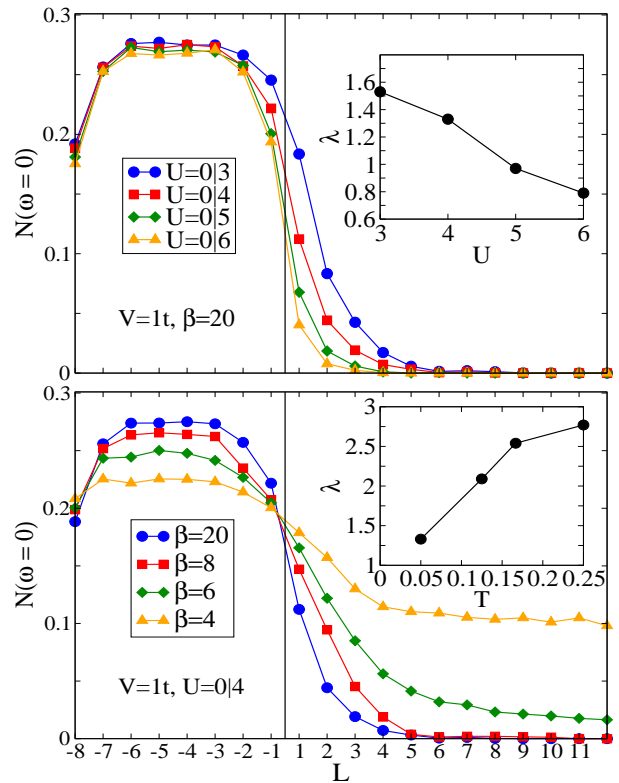

\psfig{figure=Nw0B20vsU.eps,height=5.0cm,width=8.0cm,angle=0,clip}\\
\psfig{figure=Nw0U4vsT.eps,height=5.3cm,width=8.0cm,angle=0,clip}\\
\caption{(color online) $L$-dependent density of states for different
  on-site repulsions $U$ and temperatures $T$. The smooth penetration
  of metalicity into the insulator and diminished metallic behaviors
  in the metal are clearly seen. Similar to Fig.~\ref{SafU4vsT}, we
  employ a hyperbolic fitting function and extract a penetration depth
  $\lambda$ (insets), which we find decreases with larger correlation
  $U$ and increases with higher temperature $T$.}
\label{Nw0B20vsU}
\end{figure}

\section{Density of states}

One criterion to distinguish metal from insulator is the single
particle density of states, which can be extracted from the local
imaginary-time dependent Green's function
$G_{L}(\tau)=\sum\limits_{i\sigma} T c_{iL\sigma}(\tau)
c^{\dagger}_{iL\sigma}(0)$ from
\begin{equation}
   G_{L}(\tau)= \int_{-\infty}^{\infty}\ d\omega \frac{e^{-\omega\tau}}{e^{-\beta\omega}+1}\ N_{L}(\omega)
\label{aw}
\end{equation}
and using the maximum entropy method~\cite{maxent}.
The density of states is probed experimentally with photoemission
spectroscopy.  In a translationally invariant system one often also
calculates the momentum dependent spectral function $A(k,\omega)$,
measured in angle-resolved photoemission spectroscopy (ARPES).

As mentioned in the introduction, a dramatic experimental achievement
in the past few years has been the development of standing wave ARPES
\cite{gray10} which has enabled the probing of strong correlation
effects layer by layer in a sample.  Here we make some predictions for
possible features to be seen in SWARPES by obtaining the row dependent
single particle density of states.  As with the row-by-row magnetic
correlations, $N_L(\omega)$ shown in Fig.~\ref{Axel} characterizes the
penetration of metalicity into the insulator.  Kondo proximity effects
are evident in the evolution of $N_L(\omega)$.  Figure \ref{Axel}
shows that the insulating lines for $L=6,5,4$, relatively far from the
interface, are characterized by the presence of a Kondo resonance peak
split by antiferromagnetic order at half-filling.  However as the
interface is approached for $L=3,2,1$ the gap is increasingly filled
in, evidence of the coupling to the metallic half of the lattice.
Meanwhile the metallic line ($L=-1$) immediately adjacent to the
interface shows some influence of the boundary, albeit a rather small
one: the central peak at $\omega=0$ shows a slight dip.  This is
completely gone for $L=-2$.  Evidentally the metallic behavior of
$N(\omega)$ penetrates much further into the insulator than the
insulating physics does into the metal.

Further information on the spectral weight at Fermi level
$N_{L}(U,T)|_{\omega=0}$ as a function of on-site repulsion $U$ and
temperature $T$ is shown in Fig.~\ref{Nw0B20vsU}.  The
penetration of metallic behavior into the insulator and,
conversely, the diminished
spectral weight at the Fermi surface in the metal
are clearly seen. Similar to
Fig.~\ref{SafU4vsT}, we adopt a hyperbolic function for fitting.  The
resulting penetration depths $\lambda$ are given in the inset, and
decrease with larger correlation $U$ and lower temperature $T$.

\section{Effect of variation of interface hopping}

\begin{figure}
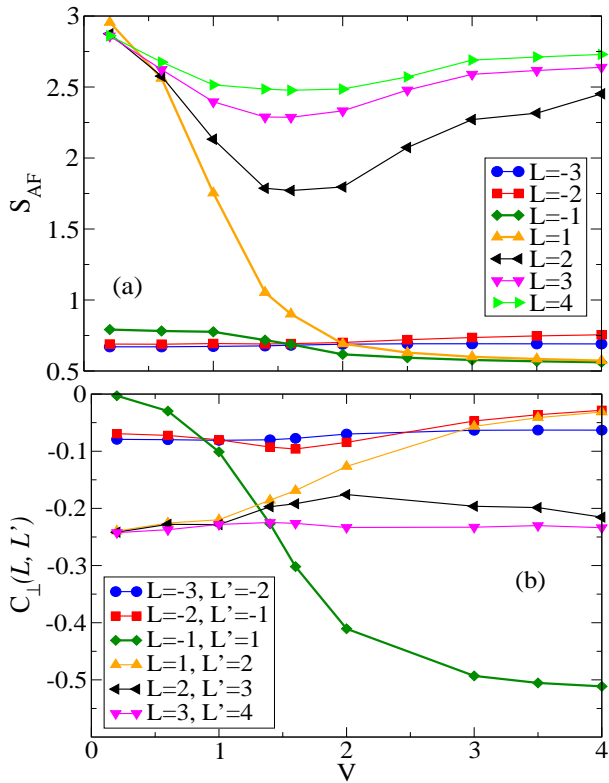

\psfig{figure=fig4z.eps,height=5.0cm,width=8.0cm,angle=0,clip}\\
\psfig{figure=fig1.eps,height=5.3cm,width=8.0cm,angle=0,clip}\\
\caption{(color online) (a) Evolution of the antiferromagnetic (AF)
  structure factors $S_{AF}(L)$ with interface hybridization $V$.
  Except the line of sites $L=1$, adjacent to the metal, $S_{AF}(L>1)$
  exhibits a nontrivial revival with increasing the $V$.  We attribute
  this affect to singlet formation at the metal-insulator interface
  which leaves the lines $L>1$ decoupled from the metal.  (b) Further
  evidence of the strong magnetic coupling across the
  interface. $U=0|4, \beta=10$. }
\label{fig4z}
\end{figure}

We show in Fig.~\ref{fig4z} the evolution of the line dependent AF
structure factors $S_{AF}(L)$ with interface hopping $V$.  As
expected, all lines $L>0$ with nonzero on-site repulsion $U=4$ have
large $S_{AF}$ when the metal and insulator are decoupled ($V =
0$). With increasing hybridization all $S_{AF}(L>0)$ initially
decrease. The most dramatic feature in Fig.~\ref{fig4z}(a) is the
different behaviors of $S_{AF}(L=1)$ and $S_{AF}(L>1)$. In the line
$L=1$, adjacent to the metal-insulator interface, $S_{AF}(L=1)$
monotonically decreases with $V$.  However, $S_{AF}(L>1)$ show
nontrivial revivals with increasing $V$ and ultimately recover to
values characteristic of the decoupled case $V = 0$ when there is no
contact with the metal whatsoever.  This behavior of $S_{AF}(L)$ is
qualitatively similar to the multi-layer metal-insulator interface in
the three-dimensional Hubbard model~\cite{euverte12},
and can be attributed to
``singlet'' formation at the metal-insulator interface, which both
suppresses $S_{AF}(L=1)$ and also leaves the remaining
lines $L>1$ decoupled from the metal.

The ``metallic'' line structure factors $S_{AF}(L)$ retain their small
values for the entire range of $V$.  As with the real space spin
correlation functions, there is a greater effect of the metal on the
insulator (suppressing magnetic order) than of the insulator on the
metal (inducing magnetic order).

Figure \ref{fig4z}(b) provides further evidence of the
strong magnetic coupling
across the interface. The dominant feature is the rapid increase of
antiferromagnetic correlation across the metal-insulator interface
$C_{\bot}(L=-1,L'=1)$ as increasing the hopping $V$. Note also that
magnetic correlation $C_{\bot}(L=1,L'=2)$ become smaller with larger
hopping $V$ due to the ``singlet'' formation at the interface $L=-1,1$.

\section{Conclusion}

We have used the numerically exact finite-temperature determinant
Quantum Monte Carlo (DQMC) method to study a metal-insulator interface
in a two-dimensional square lattice.  By investigating the long-range
antiferromagnetic order and density of states, we demonstrated that
the metallic behavior penetrates into the insulator for several lattice
spacings, with a penetration depth $\lambda$ which decreases with
increasing on-site repulsion, $U$, on the insulator side of the
interface.  $\lambda$ is also (somewhat more weakly) temperature
dependent, decreasing as $T$ is lowered towards the critical
temperature $T_c=0$ for long range magnetic order on the insulating
side.  The penetration length $\lambda$ thus shows an opposite
temperature dependence to that of the spin correlation length, $\xi$,
which grows as $T$ is lowered.

The insulator-induced antiferromagnetic long-range order in the metal
is predominantly limited to the line immediately adjacent to the
interface.  That is, magnetic characteristics of strong correlation
appear to penetrate less deeply into the metallic side of an interface
than does weak correlation physics penetrate into the insulator.  This
is consistent with previous results for a planar interface in a 3D
lattice\cite{euverte12}.  We note, however, that in the 3D geometry,
the effect of the contact with the insulator on the in-plane
conductivity $\sigma$ in the $U=0$ half of the lattice can extend
beyond the contact layer.  This is consistent with our results for the
density of states at the Fermi level $N(\omega=0)$, which are shifted
from their $U=0$ values for layer $L=-2$ in addition to layer $L=-1$.

In the past several years, studies of the effect of spatially varying
densities and interaction strengths have been motivated by experiments
on trapped atomic gases\cite{jaksch98,batrouni02,stoferle06,chin06}.
In such systems the spatial variation, eg a quadratic confining
potential $V_{\rm trap}(r) = V_{T} (r/l)^{2}$, has an explicit length
scale $l$.
This complicates the determination of intrinsic length scales
associated with the response of the interacting fermions themselves.
Our choice here of a sharp (scale free) interface between metal and
insulator ($U_{L} = 0, L=-1,-2,-3...$ and $U_{L} = U>0, L=1,2,3...$)
allows us to attribute the lengths characterizing the relaxation of
properties on either side of the interface solely to the fermionic
correlations.  This choice is, of course, also more appropriate to the
solid state context of a sharp interface between two materials.

One further motivation to study the metal-insulator interface is its
close relation to the question of ``orbitally selective Mott
transitions''
(OSMT)\cite{liebsch04,arita05,koga04,ferrero05,demedici05,knecht05}.
Here the central question is whether orbitals with different degree of
electronic correlation, coupled together by interorbital
hybridization, necessarily undergo the Mott transition simultaneously.
The layer index `L' in the Hamiltonian considered in this paper bears
a formal similarity to the orbital degree of freedom in the OSMT,
although of course the details of the coupling via hybridization are
rather different in the two cases.  It is evident from our data that a
layer which shows the hallmarks of an antiferromagnetic Mott insulator
can coexist with layers which have the characteristic behavior of a
paramagnetic metal.  That such coexistence is possible is similar to
the conclusion ultimately reached in numerical studies of the OSMT.

In this paper, the density of states and magnetic correlations near an
interface between $U=0$ and $U \neq 0$ regions have been found to be
more or less smooth interpolations between the bulk metal and Mott
insulator.  On the other hand, Quantum Monte Carlo simulations have
observed novel phases\cite{novel} such as spin liquids, to arise in models where the
energy scales are poised at the boundary between a semi-metal and an
antiferromagnet.  An interesting extension of the present work would be,
therefore, to bring such more general regions into contact and study the
properties at the interface.

\section{Acknowledgements}
We acknowledge support from the National Scence Foundation under grant
NSF-PIF-1005502.  This work was also supported under ARO Award
W911NF0710576 with funds from the DARPA OLE Program and by the CNRS-UC
Davis EPOCAL LIA joint research grant.  We are grateful to Bo Deans
for useful input.



\begin{thebibliography}{40}
\bibitem{footnote3}
The linear scaling in $\beta$ is restricted to cases when the matrix
condition number does not become overly large:  At large interaction
strengths $U$ the necessity of recomputing the green's function from
scratch due to numerical instabilities
can introduce $\beta^2$ terms.  These can also arise in
computation of the imaginary time dependent Greens function.
\bibitem{footnote1}  `Long range order' might be a bit of a misnomer.
Although the correlations remain non-zero at long distances, it
is likely this reflects an imposed order from the external
contact with the insulator,
as opposed to order mediated through the metallic layer itself.
\end{thebibliography}
\end{document}